\documentclass[b5paper,twoside]{jpconf}
\usepackage{graphicx}

\begin{document}
\title{ New Zealand pathway towards Asia-Pacific and global e-VLBI research and development }

\author{S Gulyaev$^1$,T Natusch$^1$,S Weston$^1$ and P Thomasson$^{1,2}$}
\address{$^1$Institute for Radio Astronomy and Space Research, Auckland University of Technology, Private Bag 92006, Auckland, New Zealand}
\address{$^2$The University of Manchester, Jodrell Bank Observatory, Macclesfield, Cheshire SK11 9DL, U.K.}
\ead{Sergei Gulyaev <sergei.gulyaev@aut.ac.nz>}

\begin{abstract}

Over the past 3 years, Auckland University of Technology has established the first radio astronomical observatory in New Zealand, which, because of its remote geographic location, has quickly become a member of a number of international VLBI networks, in particular the IVS and the LBA.   Not only has this added significantly to the observational power in the Pacific and Oceania, but by utilising new fibre connections within New Zealand, and across the Pacific and the Tasman Sea, the New Zealand radio telescopes have now been linked to many in Australia, Asia and the Pacific.   Recent astronomical results are presented and plans for widening New Zealand participation in Australasian, Asia-Pacific and global VLBI research and development are outlined.  Real-time e-VLBI is a vital part of New Zealand's capability development towards the SKA.  The rapid and challenging establishment of New Zealand radio astronomy can serve as a model for the engagement in mega-Science and e-Science by resource-limited institutions and nations.   Perspectives for collaboration between New Zealand and Thailand in the field of radio astronomy are included.

\end{abstract}

\section{Introduction}
\begin{figure*}
\begin{center}
\includegraphics{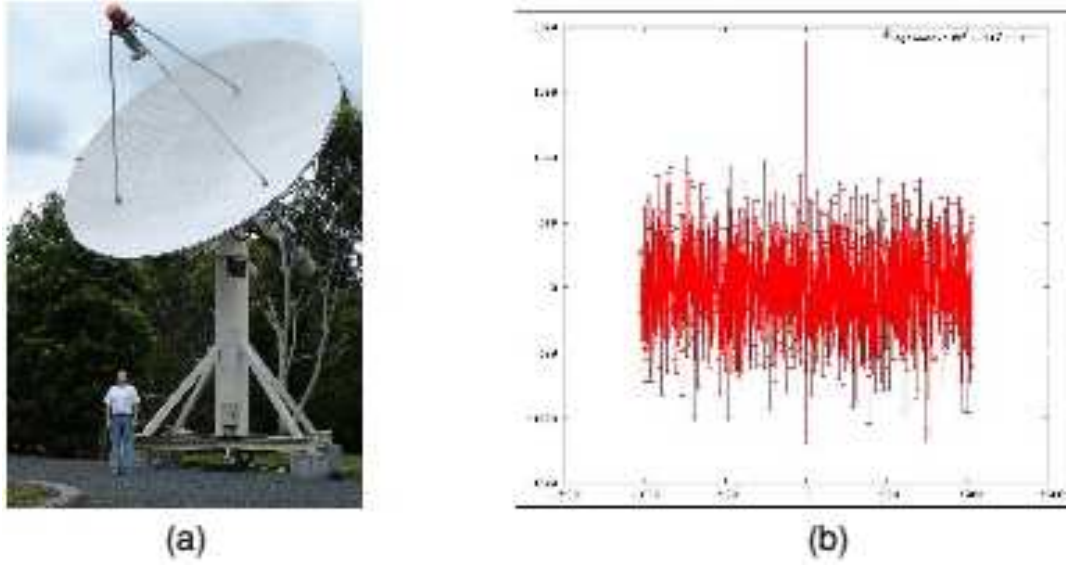}
\caption{(a) The Karaka 6-m telescope  (b) First fringe from PKS\,1921-231}
\label{x6m}
\end{center}
\end{figure*}

     The first radio astronomy undertaken in New Zealand was in the very early pioneering days in 1948.  John Bolton and Gordon Stanley, from CSIRO in Australia, initially set up a cliff-top interferometer using the sea as a mirror, first near Pakiri, $\sim80\,km$ north of Auckland on the east coast of North Island, and then later at a World War II radar station near Piha, to the west of Auckland.  The interferometer operated at 100\,MHz, and Bolton and Stanley were able to locate for the first time the sources of radio waves from three sources, Taurus\,A, Centaurus\,A and Virgo\,A, which were now seen to be outside the solar system.  This was probably the opening of a new window on the Universe, but it was not until 2005 that further significant radio astronomy was conducted in New Zealand. Brent Addis, a radio `ham', had built a 6-m radio telescope at Karaka (Figure~\ref{x6m}(a)), to the south of Auckland, and this was used at 1.6\,GHz with the Australia Telescope Compact Array at Narrabri (ATCA) for the first VLBI observations across the Tasman Sea by the newly formed group at Auckland University of Technology (AUT) and scientists at the Australia Telescope National Facility (ATNF).

\begin{figure*}
\begin{center}
\includegraphics[scale=0.5]{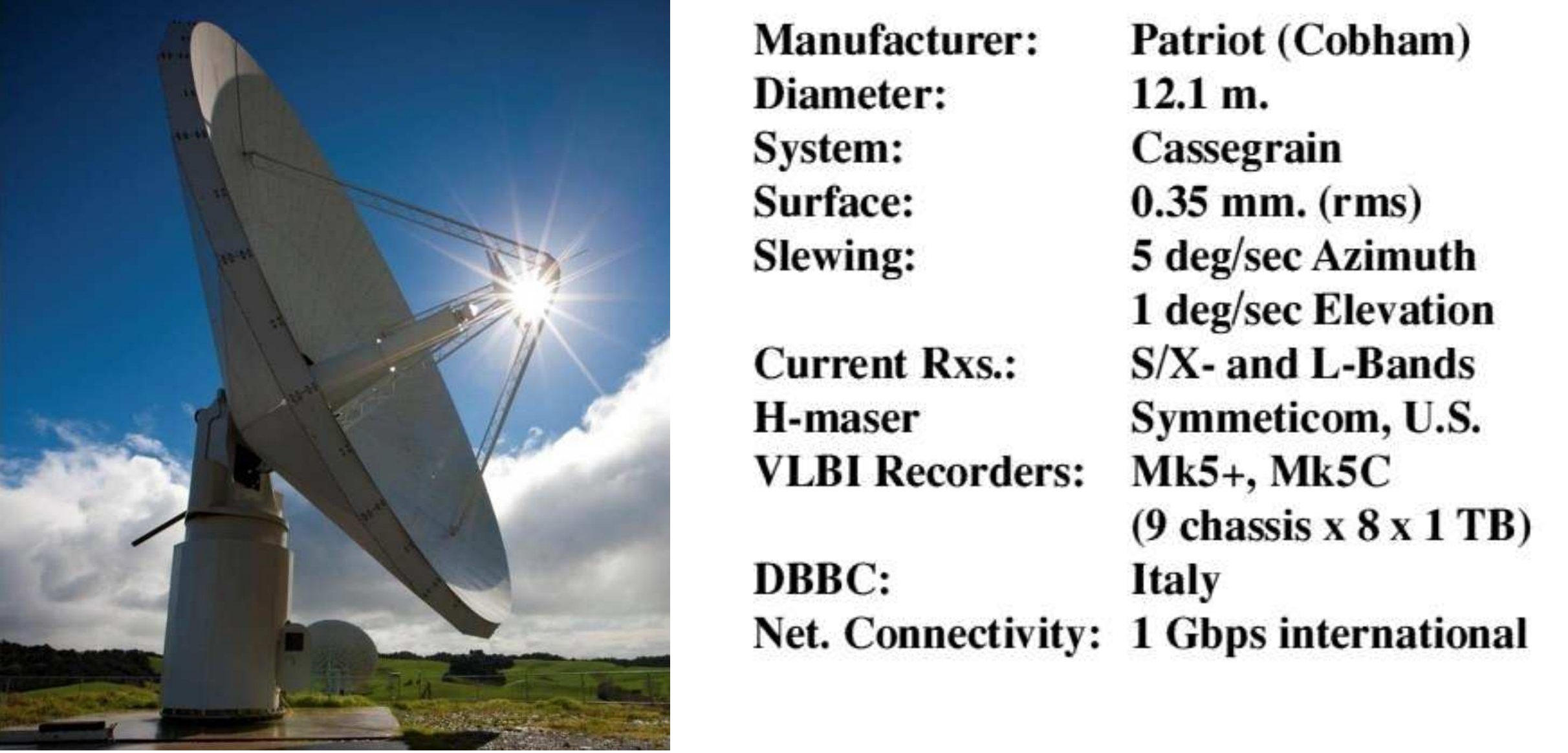}
\caption{
The 12-m Patriot Antenna and its parameters}
\label{Pat-12m}
\end{center}
\end{figure*}

\begin{figure*}
\begin{center}
\includegraphics[scale=0.7]{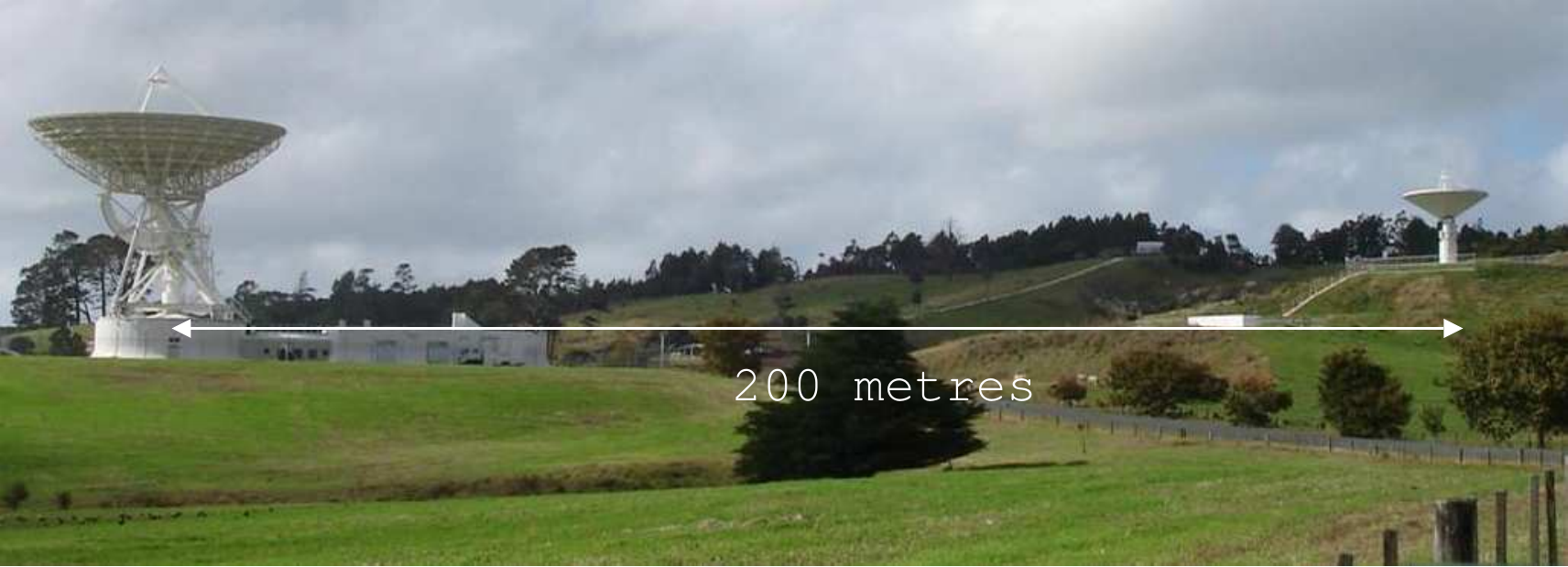}
\caption{
The Warkworth Observatory showing the relative positions of the 2 telescopes}
\label{warkobs}
\end{center}
\end{figure*}

Figure~\ref{x6m}\,(b) shows the first fringe from the radio telescope at 1.6 GHz.  Since then, AUT has acquired a 12.1-m Cassegrain radio telescope manufactured by Patriot (See Figure~\ref{Pat-12m} for its parameters and associated equipment), which has been sited in a radio-quiet zone (shown to be radio quiet by actual measurements) $\sim55\,km$ to the north of Auckland and a few kilometres to the south of Warkworth.  
It was officially opened on 8\,October\,2008. More recently, a former satellite-communications 30.5-m\,radio telescope on the same Warkworth site, and $\sim$200m from the 12-m antenna, has been taken over by AUT. This telescope, currently being upgraded, is capable of operations up to 40\,GHz. So the Warkworth Observatory has come into being (Figure \ref{warkobs}).

\section{Network Connectivity}
The introduction of the Kiwi Advanced Research and Education Network (KAREN) in 2005 and the linking into it of the Warkworth Observatory somewhat later has enabled data to be transferred between the Warkworth Observatory and Australia and the U.S.A, and onwards further to Europe and Japan at data rates of 1�Gbps.  Initial VLBI test observations between the 12-m\, telescope and Australian Long Baseline Array (LBA)  telescopes in Australia have shown that New Zealand can contribute on a regular basis to LBA observations and a formal joining with the LBA was signed in 2010.

\section{Australasian SKA and Trans-Tasman Collaboration} 
\begin{figure*}
\begin{center}
\includegraphics[scale=0.5]{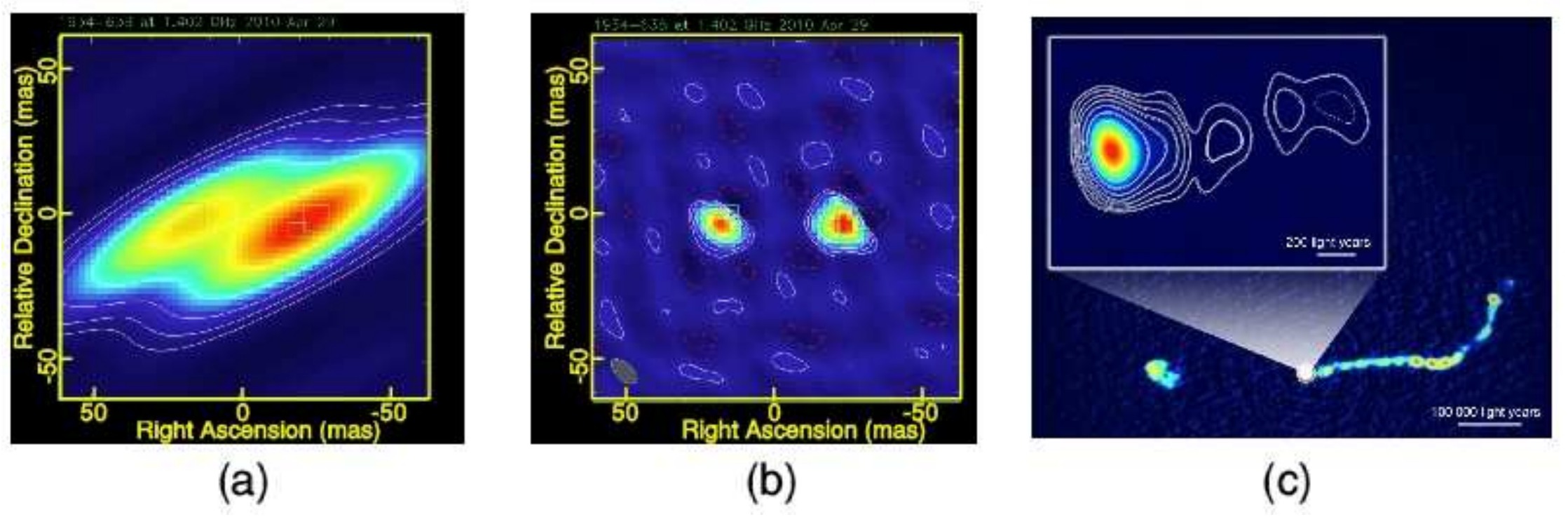}
\caption{(a)\,PKS 1934-638 at 1.4 GHz ('normal' LBA)
\\(b)\,PKS\,1934-638 at 1.4 GHz (LBA+ASKAP+Warkworth)
\\(c)\,PKS\,0632-752 at 1.4 GHz (LBA+ASKAP+Warkworth) Credit:Steven\,Tingay}
\label{sources}
\end{center}
\end{figure*}

The introduction of the Australian Square Kilometre Array (SKA) Pathfinder telescopes (ASKAP) at the Murchison Radio Observatory in Western Australia and also the inclusion of the Warkworth 12-m telescope, has meant that the maximum baseline of the LBA has been extended to $\sim5500\,km$ with a consequential improvement in both resolution and uv$�$coverage. Figures \ref{sources}(a) and (b) show the results of observations of the Gigahertz Peaked Spectrum radio galaxy, PKS 1934-638, at 1.4 GHz with and without the ASKAP and Warkworth telescopes.  It is interesting to note that were there to be a telescope in northern Thailand, the maximum length of baseline from an ASKAP antenna in Western Australia to this would be $\sim4400\,km$, but in a much more NS direction.  The uv-coverage of such an enhanced LBA array would improve significantly, not only from the point of view of the intermediate baseline lengths that would be generated, but also because of the considerably improved NS baselines.

For the above VLBI observations, data have been recorded on disc and correlated at a later time.  For the successful operation of the SKA, `real-time' correlation is essential, and this was achieved including the Warkworth 12-m telescope on 1st\,February 2011, when a real-time e-VLBI fringe at 1.4\,GHz between it and the 22-m Mopra telescope in Australia was obtained.  Somewhat later (June 2011), the first `real-time' observations of the quasar, PKS 0637-752, using all of the LBA telescopes, including the ASKAP and Warkworth dishes were successfully made.  Figure \ref{sources}(c) shows the resulting image.     

\section{The International VLBI Service for Geodesy and Astrometry (IVS)}
It has been long established that VLBI observations can contribute considerably to the study of the Earth's crust and its movements, and a network of telescopes across the world has been established for these observations.  The inclusion of a telescope in New Zealand, which is located in a very active region of the Earth's surface, should be able to provide a significant contribution to the network.  Following successful test VLBI observations at both S-Band and X-Band frequencies using the Warkworth 12-m telescope and a 32-m telescope at Tsukuba in Japan, the Warkworth 12-m\,telescope has now taken part on a regular basis in several observing sessions of this network since February 2011.  It is also participating in the AuScope Project, an Australian geodetic research programme.

\section{Conclusion}
The observations showing the development of radio astronomy in New Zealand have clearly shown that:- 
\\ New Zealand has a strong observational basis for VLBI
\\ New Zealand is capable of electronically transferring large amounts of data
\\ New Zealand has excellent radio-quiet zones.

\ack
This research was supported by grants from the New Zealand Ministry for Economic Development.  Warkworth connectivity via KAREN was made possible by the Remote Site Connectivity Fund provided by the Ministry of Research Science and Technology (MoRST) on behalf of the New Zealand Government. We are grateful to IBM for providing the IRASR with the IBM Blade Centre. Thanks also for their support and assistance go to staff from KAREN and AUT's ICT Department.  Our special thanks go to our VLBI colleagues in Australia at ATNF, the International Centre of Radio Astronomy Research (in Western Australia) and at the University of Tasmania.
\end{document}